\documentclass[9.2 pt, conference]{IEEEtran}

\IEEEoverridecommandlockouts
\IEEEpubid{\makebox[\columnwidth]{This work was supported by funding from NIDA (R34DA050256) \hfill} \hspace{\columnsep}\makebox[\columnwidth]{ }}
\usepackage{cite}
\usepackage{amsmath,amssymb,amsfonts}
\usepackage{algorithmic}
\usepackage{graphicx}
\usepackage{textcomp}
\usepackage{xcolor}
\usepackage{subfiles}
\usepackage[linesnumbered,ruled]{algorithm2e}
\SetAlFnt{\small}
\usepackage{algorithmic}
\algsetup{linenosize=\tiny}
\usepackage[color=white]{todonotes}
\usepackage{subcaption}
\usepackage{multirow}
\usepackage{url}
\usepackage[utf8]{inputenc}
\usepackage[export]{adjustbox}

\usepackage{booktabs}
\usepackage{tabularx}
\makeatletter
\let\NAT@parse\undefined
\makeatother
\usepackage{hyperref}

\usepackage{authblk}
\author[1]{Mohammad Nur Hossain Khan}
\author[2]{Jialu Li}
\author[3, 4]{Nancy L. McElwain}
\author[2, 4]{Mark Hasegawa-Johnson}
\author[1]{Bashima Islam}
\affil[1]{Department of Electrical and Computer Engineering, Worcester Polytechnic Institute}
\affil[2]{Department of Electrical and Computer Engineering, University of Illinois}
\affil[3]{Department of Human Development and Family Studies, University of Illinois}
\affil[4]{Beckman Institute for Advanced Science and Technology, University of Illinois}
\affil[  ]{mkhan@wpi.edu, jialuli3@illinois.edu, mcelwn@illinois.edu, jhasegaw@illinois.edu, bislam@wpi.edu}
\begin{document}

%\title{Deadline Aware Intermittent System by Scheduling Computing and Harvesting Tasks}

\title{Sound Tagging in Infant-centric Home Soundscapes \vspace{-2ex}}

\maketitle

\begin{abstract}
Certain environmental noises have been associated with negative developmental outcomes for infants and young children. Though classifying or tagging sound events in a domestic environment is an active research area, previous studies focused on data collected from a non-stationary microphone placed in the environment or from the perspective of adults. Further, many of these works ignore infants or young children in the environment or have data collected from only a single family where noise from the fixed sound source can be moderate at the infant's position or vice versa. Thus, despite the recent success of large pre-trained models for noise event detection, the performance of these models on infant-centric noise soundscapes in the home is yet to be explored. To bridge this gap, we have collected and labeled noises in home soundscapes from 22 families in an unobtrusive manner, where the data are collected through an infant-worn recording device. In this paper, we explore the performance of a large pre-trained model (Audio Spectrogram Transformer [AST]) on our noise-conditioned infant-centric environmental data as well as publicly available home environmental datasets. Utilizing different training strategies such as resampling,  utilizing public datasets, mixing public and infant-centric training sets, and data augmentation using noise and masking, we evaluate the performance of a large pre-trained model on sparse and imbalanced infant-centric data. Our results show that fine-tuning the large pre-trained model by combining our collected dataset with public datasets increases the F1-score from 0.11 (public datasets) and 0.76 (collected datasets) to 0.84 (combined datasets) and Cohen's Kappa from 0.013 (public datasets) and 0.77 (collected datasets) to 0.83 (combined datasets) compared to only training with public or collected datasets, respectively. 
    % Classifying and detecting sound events in domestic environments is an active research area. To accomplish this task, the previous study used data collected from devices with a static position and from the perspective of parents. It is imperative to collect data from an infant's perspective to consider the added noise and movement of microphones when analyzing the data. Recently, sound event detection with large pretrained models has been shown to have state-of-the-art performance. However, the performance of these models on infant ego-centric noise soundscape is still unexplored. In this work, we collected and labeled home environmental sounds from an infant's perspective using a multimodal recording device. We explored the performance of a large pretrained model (Audio spectrogram transformer) on this noise-conditioned infant-egocentric environmental data. Our results show that the use of publicly available noise data along with our own data achieves higher and more reliable performance.
\end{abstract}
\vspace{-1ex}
\section{Introduction}

% Mental health problems starting in early childhood are often neglected in the US due to the lack of continuous monitoring and observation. Though regular hospital visits have resulted in adequate diagnosis of developmental disorders after the age of 3, very little focus has been given to children under 3 years old ~\cite{bitsko2016health}. Researchers have recently looked into identifying signs of distress in infants and young children ~\cite{mcelwain2006maternal} as early detection of development disorders opens the door to early intervention.

A host of studies indicate that certain environmental noises may have negative health and psychological outcomes, including but not limited to elevated blood pressure, elevated endocrine response, sleep disturbance, poor cardiovascular functioning, mental health disturbance, and decreased cognitive functioning for both young children \cite{stansfeld2015health} and adults \cite{li2022environmental}. Additionally, both human and animal studies underscore the deleterious physiological and biological effects of environmental noise in utero and early infancy \cite{stansfeld2015health}, and indicators of household or environmental noise have been associated with decreases in attention~\cite{werchan2022signal}, and speech perception/language learning~\cite{erickson2017influences} during the first years of life. Mechanisms through which noise may adversely affect child outcomes are likely to be direct via dysregulated stress physiology~\cite{wass2019influences} or indirect via adults' annoyance or irritability~\cite{li2022environmental} due to environmental noise. Further, certain types of environmental noise (i.e., intermittent or unpredictable nonlinguistic noise) pose greater developmental risks compared with other noise types, including more predictable noise (e.g., white noise)~\cite{erickson2017influences, werchan2022signal}. Thus, in this work, we aim to detect the presence of different types of household noise (white noise, adult speech, TV, percussive noise, music, child voice, and background noise) that have the potential to provide novel insights into the effects of noise on infants' physiological and behavioral health.

%Though acoustic researchers widely explore sound classification in household scenarios \cite{foster2015chime, piczak2015esc, turpault2019sound}, they are collected from the perspective of a global static position in the environment ~\cite{foster2015chime, piczak2015esc}. 
Prior work most closely resembling the current study has used the Language ENvironment Analysis (LENA) system \cite{xu2009reliability}, which includes an audio recorder worn by the child in the home environment and proprietary software that automates the detection of classes of interest, including child vocalizations, adult speech, electronics, and overlapping speech/noise. Beyond a LENA technical report \cite{xu2009reliability}, a limited number of studies have assessed the performance of the LENA algorithm and indicate somewhat low performance when correcting for chance agreement (Cohen’s kappa=0.28) and wide variability in F1 scores for the four key classes: child = 0.37, adult =0.85, electronics =0.49, and overlap = 0.05 \cite{bulgarelli2020look}. Using daylong LENA recordings collected among 22 infants, Khante et al. \cite{khante5auditory} applied novel algorithms to detect levels of household auditory chaos (4 classes: 1=no chaos to 4=high chaos) and achieved the best performance using a Convolutional Neural Network (CNN) on 40 hours of balanced annotations (Macro F1 = 0.701). Although novel, the classification of household chaos is agnostic regarding specific noise types and thus may lose important information relevant to child functioning. Taken together, this prior work underscores the challenges of tagging sounds in infant-centric home recordings and also indicates the need for the current work, which will classify a wider range of household sounds – chosen for their developmental significance - than has been previously attempted. 

Acoustic context monitoring from an infant's perspective introduces unique challenges -- (1) the position of the microphone is mobile as the infant wearing it constantly changes their position and location; (2) the intensity of the audio stochastically changes due to the infant's changing proximity to the sound sources and the presence of additional obstacles, e.g., when the parent carries the baby; (3) unlike existing works on home environmental noise classification, a baby-worn microphone data is polluted by additional noises such as the baby's own vocalizations, and (4) the lack of existing labeled datasets recorded in home environments without the presence of on-site annotator make developing such an automated sound tagging system complicated.

In this study, we use an infant wearable multi-modal device called LittleBeats$^{\text{TM}}$ (LB) that has been utilized in prior research \cite{li23e_interspeech} on automated speaker diarization (SD) and vocalization classification (VC) for infants and parents in the home environment. The LB device is housed in the chest pocket of a specially design infant shirt and continuously collects audio from the infant's perspective in the home environment.
In this work, we develop an automatic sound detection pipeline to classify noise collected in an infant-centric soundscape (i.e., sounds recorded from devices worn by the infant and thus from the infant's perspective). We collected and labeled 3.91 hours of audio data using LB devices from 22 families with children under 14 months of age. 

%Our closest work uses the Language ENvironment Analysis (LENA) systems \cite{greenwood2011assessing}, where the infant wears an audio recorder in the home environment~\cite{khante2019quantifying}. It used unsupervised clustering based on statistical and learned features to identify chaos and only evaluated nine infants. Their hierarchical unsupervised clustering classifies acoustic sounds to identify chaos (indicated by the baby crying and human sound), which only focuses on limited classes. Some study uses LENA for TV and voice recognition~\cite{gilkerson2017evaluation}.  

To address the above-mentioned challenges and to provide robust sound tagging in infant-centric noise soundscapes in the home, we first explore the potential of audio representation from a pre-trained model, Audio spectrogram transformer (AST)~\cite{gong21b_interspeech}, trained on a large public dataset, AudioSet~\cite{gemmeke2017audio}. Next, given our limited LB training data for training a whole model, we demonstrate that using a pre-trained model and data from a public dataset to fine-tune for the downstream sound classification task can be beneficial. Finally, we evaluate our algorithm on our LB home data and public datasets. To the best of our knowledge, this is the first study to develop an environmental sound classification pipeline on data collected from all relevant noise sources from an infant's perspective.

%To our knowledge, this is the first study to focus on continuously or passively monitoring a baby's acoustic environment that considers all relevant noise sources.

% Researchers or parents have previously used a cell phone, video camera, or 

% \note{add some text}. 

% For sound classifications, researchers collected clean audio in home soundscape[chime] or individual household appliance sounds that ignored human or child vocalization [esc-50, desed]. To our knowledge, no home soundscape dataset is collected from the perspective of an infant. Getting the data from an infant's perspective is crucial for several reasons. An infant can change position very frequently and constantly move, which will change the microphone's position and add different intensity levels for different audio segments. Additionally, collecting data from the infant's perspective will create different noises such as the rubbing of the clothes or rustling noise all the time, which will make the detection task difficult. 

\section{Background and Related Work for Sound Event Detection}
Various acoustic features have been used for sound event detection, including Mel scaled spectrogram~\cite{gong21b_interspeech}, Mel-Frequency Cepstral Coefficients (MFCC)~\cite{piczak2015esc}, and log-power spectrogram ~\cite{guzhov2021esresnet} and even raw waveform~\cite{tokozume2017learning, zhu2018environmental}. Convolutional neural networks alone~\cite{tokozume2017learning}, at multiple time scales~\cite{zhu2018environmental}, and with gamma tone filterbanks~\cite{zhu2018environmental} have been used to extract significant features from raw acoustic data. Some studies~\cite{guzhov2021esresnet, gong21b_interspeech} have used weight initialization from popular pre-trained vision models, e.g., DeiT~\cite{touvron2021training} to improve performance. Recently Audio spectrogram transformer~\cite{gong21b_interspeech} and WHISPER-AT~\cite{gong_whisperat} used a transformer-based model with initial weights from ImageNet~\cite{deng2009imagenet} and WHISPER~\cite{radford2023robust} respectively. These models are trained with large datasets AudioSet~\cite{gemmeke2017audio} and WHISPER~\cite{radford2023robust}, which are 4971 hours and 680,000 hours long, respectively. Due to the exposure to such large datasets, these models are more robust on unseen data that are not used to train these models. However, their performance in a real-world environment scenario or collected by an infant-worn microphone, such as a home, has yet to be explored.

\section{Data}
\label{sec:data}

We assess the performance of a large pre-trained model on household noises using public datasets and collected data at home using infant-worn LB devices (see~\cite{nancy2023} for more details about the device setup and home use). To characterize the infant-centric soundscape, we divide the audio segments into seven categories --  child voice, adult speech, the sound of television, percussive noise, white noise or silence, music, and background noise (household appliance). Table~\ref{tab:data_distribution} shows the distribution of data from the public datasets and collected using LB device.

\begin{table}[t]
\caption{Data distribution (in minutes) of different classes.}
\label{tab:data_distribution}
% \vspace{-1em}
\resizebox{0.49\textwidth}{!}{
\begin{tabular}{@{}cccccc@{}}
\toprule
                    & \multicolumn{4}{c}{Public data}        & \multicolumn{1}{c}{\multirow{2}{*}{LB home audio}} \\ \cmidrule(r){1-5}
                    & CHiME-home & ESC-24 & GTZAN & Libritts & \multicolumn{1}{c}{}                               \\ \cmidrule(l){1-6} 
Child voice         & 79.8       & -      & -     & -        & 53.4                                                \\
Adult speech        & 45.0        & -      & -     & 53.4     & 53.4                                                \\
TV                  & 78.6       & -      & -     & -        & 21.0                                                \\
Percussive noise    & 50.4        & 31.8    & -     & -        & 34.2                                                \\
White noise         & 1.8         &        & -     & -        & 53.4                                                \\
Music               & -          & -      & 53.4   & -        & 19.2                                                \\
Household appliance & 3.0         & 31.8    & -     & -        & 51.0                                                  \\ \bottomrule
\end{tabular}}
\vspace{-2em}
\end{table}

\subsection{Public dataset}
We used noise data from two public datasets -- CHiME-home~\cite{foster2015chime} and ESC-50~\cite{piczak2015esc}. CHiME-home had a total of 1946 4-second long audio segments from one family. Each audio segment had one or more classes: silence, child voice, male voice, female voice, appliance noise, percussive noise, TV, other, and unknown. We discarded the other and unknown classes and merged male and female voices into adult speech. ESC-50 had 40 5-second audio clips from 50 environmental sound classes (collected from user-uploaded audio~\cite{font2013freesound}), and 24 of them were domestic sounds, where segments fall into percussive noise and household appliance sounds. We evaluated model performance on ESC-50 and only 24 classes of ESC-50 (ESC-24) that can occur in a home environment. We also re-recorded noise data from two public datasets -- CHiME-home~\cite{foster2015chime}, ESC-50~\cite{piczak2015esc} -- using the LB device in an anechoic chamber to assess whether performance varies as a function of recording set up.

To address the lack of adult speech and music in these datasets while fine-tuning our pretrained model, we used speech and music data from a small LibriSpeech corpus (libriTTS)\cite{zen19_interspeech} and GTZAN~\cite{1021072} to balance the training/fine-tuning dataset. We collected 800 adult speech and 800 music samples for fine-tuning; each sample was 4 seconds long. Note that we did not re-record GTZAN and libriTTS audio, as these were used only for pre-training and not for evaluation. 
% we have found that re-recording does not significantly change performance. 
% \note{collected from user-uploaded audio in freesound project~\cite{font2013freesound}}; 

% \begin{figure}
%     \centering
%     \includegraphics[width = .5\textwidth]{figure/lb_data_collection.pdf}
%     \vspace{-2em}
%     \caption{Data collection using LittleBeats device~\cite{nancy2023}.}
%     \label{fig: Data Collection}
%     \vspace{-1.8em}
% \end{figure}

\subsection{Collected Infant-Centric Audio (LB Home Audio)}
Twenty-two families with infants between 0-14 months were recruited for this study through study brochures posted in local community organizations (e.g., libraries) and online forums serving families with young children. The Institutional Review Board (IRB) approved all study procedures at the University of Illinois, Urbana-Champaign. To protect participants' privacy and confidentiality of the data and to increase participants' trust, consent forms specified that identifiable information, including audio, would only be accessible to the research team. Our consent forms stated that human coders will only hear small samples of the data (the labeled data) and that the majority of the recordings will be analyzed automatically without human intervention. 
% These study procedures and policies helped increase participants' trust that the research team prioritizes participant privacy and data security/confidentiality.

To collect the data, we placed the device on the infant's chest pocket and collected daylong data (8-10 hours) from each device. We separated each daylong recording into 10-minute segments to manually annotate the collected home recordings. As continuous manual annotation of the audio recordings is time- and labor-intensive, human coders only annotated a few 10-minute segments for each family, selected based on the highest active vocalization rates computed by a statistical voice activity detector (VAD) ~\cite{sohn1999statistical}. Human coders manually labeled child, female adult, male adult, music, percussive or sharp noise, white noise, and TV sounds using Praat~\cite{boersma2007praat} (an annotation software), with cross-coder validation at a precision of 0.2s. Ten percent of selected 10-mins segments were double coded, and inter-coder reliability (Cohen’s kappa score) was between 0.80 and 0.89 for child and adult speakers. All other segments were single-coded. In total, we obtained 3.91 hours of data from 22 families.

However, we found only two background noise samples from these families. Thus, we collected background noise data using LB devices from different household appliances, e.g., seven vacuum cleaners, two washing machines, and one dishwasher in three homes. These data were collected from a static position instead of an infant's perspective.
\vspace{-1ex}
\subsection{Data Pre-Processing}
% \vspace{-1ex}
We resampled all collected data to 16kHz using librosa~\cite{mcfee2015librosa} as most pre-trained models are developed for 16KHz audio. %Deleted by Jialu: To prepare labeled data for fine-tuning, we labeled the audio stream in intervals of 4s, starting every 0.2s. The label of each 4s interval was determined by the temporal majority of human annotations on the centered 1s interval (timestamps 0.5-1.5s), if two or more audio labels were present. If only one \note{audio} label was present, its label was applied to the whole interval if its duration was 0.2s or greater, as we observed that children tend to make short vocalizations. Intervals labeled with more than one speaker were discarded. 
To prepare labeled data for fine-tuning, we extracted each labeled segment in intervals of 4 seconds. For segments shorter than 4 seconds, we appended the neighbored left and right audio contexts evenly to make up to 4 seconds.
%There was abundant child voice, adult speech, and white noise compared to other types of sound. 
For our task, we used a total of 800 segments of white noise, 318 segments of TV, 800 samples of child voice, 800 samples of adult speech, 290 samples of music, 768 samples of background noise, and 509 samples of percussive noise. We randomly split the dataset for fine-tuning (80\%) and testing (20\%), where we include non-overlapping intervals from each of our 22 families in both fine-tuning and test sets. Thus, our results are multi-family internal validation results rather than external validation results.

% For sound event detection, most studies are done on data collected in one environment or from one family ~\cite{foster2015chime, piczak2015esc}. To be consistent with other sound event detection tasks, we include non-overlapping intervals from each of our 22 families in both fine-tuning and test sets, thus our results are multi-family internal validation results rather than external validation results~\cite{gong_whisperat, gong21b_interspeech, guzhov2021esresnet}.
% \vspace{-0.2em}
\vspace{-1ex}
\subsection{Data augmentation}
% \vspace{-1ex}
We use two data augmentation techniques -- spectrogram augmentation~\cite{wei2020comparison} that incorporates frequency masking and time masking, and (2) random noise addition~\cite{wei2020comparison}.
% Data augmentation has been proven to improve performance for various tasks ~\cite{wei2020comparison}. We use two data augmentation techniques -- spectrogram augmentation~\cite{wei2020comparison} that incorporates frequency masking and time masking, and (2) random noise addition~\cite{wei2020comparison}. 
The maximum frequency and time mask lengths used in this study are 24 and 96, respectively. We further experimented with Specmixup~\cite{wei2020comparison}, where two data samples are mixed by applying time-frequency masks. Although spectrogram augmentation and random noise addition improve the model performance, we found that AST performs better without using any mixup as AST is already trained to recognize a large set of classes.

\begin{figure}[!htb]
% \vspace{-0.6em}
    \centering
    \includegraphics[width = .48\textwidth]{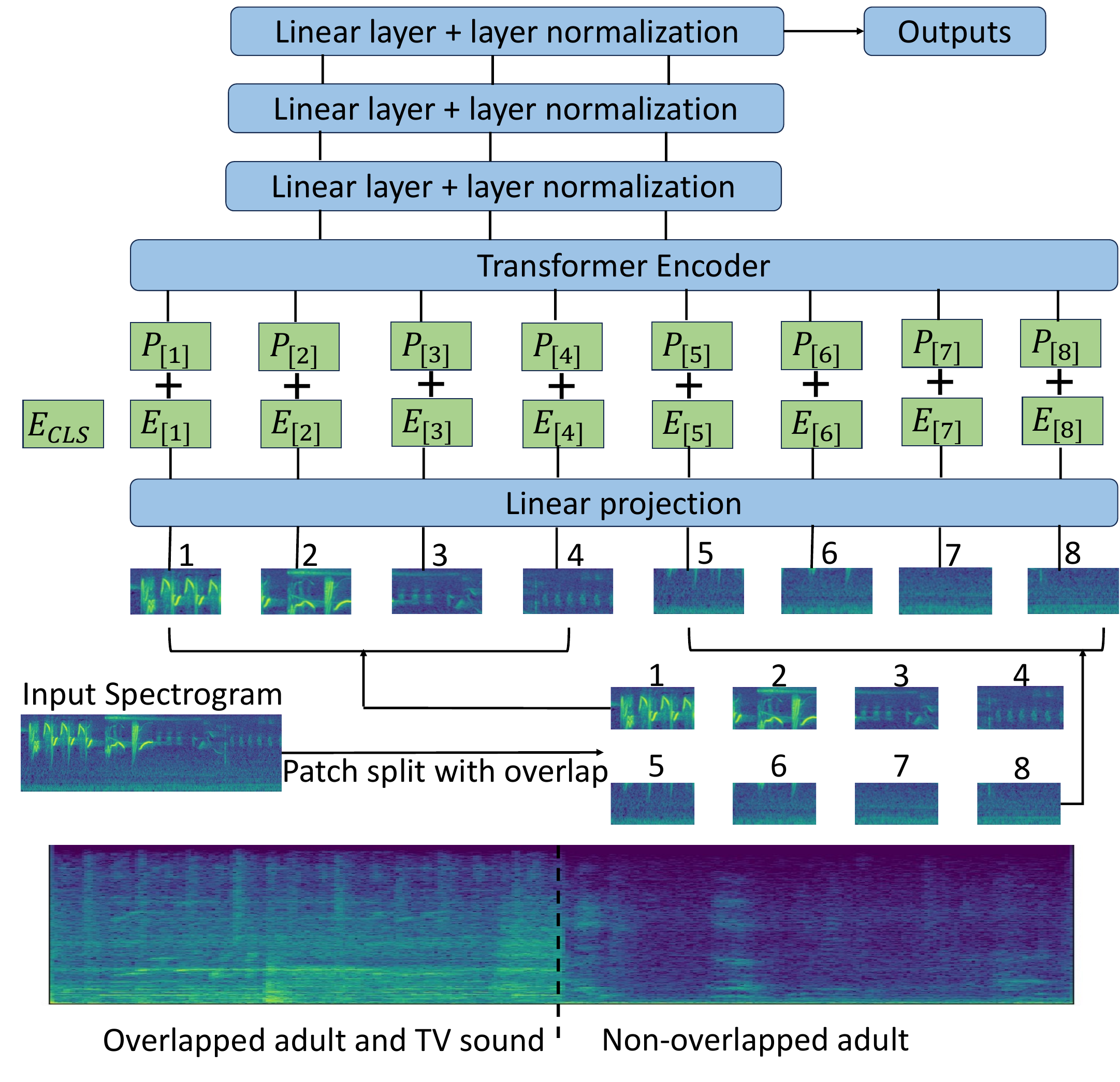}
    \vspace{-1em}
    \caption{Model architecture for fine-tuning and example of an overlapped and non-overlapped audio segments.}
    \label{fig:model architecture}
    \vspace{-1em}
\end{figure}

% \begin{figure}[!htb]
% \begin{subfigure}{0.5\textwidth}
% \centering
%       \includegraphics[width =\textwidth]{figure/model_ast_1.pdf}
%       % \caption{}
%     % % \label{fig:model architecture}
% \end{subfigure}
% % \vspace{1em}
% \begin{subfigure}{0.5\textwidth}
% \centering
%     \includegraphics[width =\textwidth]{figure/overlapped_spec.pdf}
%     % \caption{Example of an overlapped and non-overlapped audio segments}
%     % \label{fig: overlapped_spec}
% \end{subfigure}
% \caption{Model architecture used to fine-tune in the proposed study and example of an overlapped and non-overlapped audio segments}
% \label{fig: Model architecture}
% \end{figure}
\section{Experimental setup}
\vspace{-1ex}
Audio Spectrogram Transformer (AST)~\cite{gong21b_interspeech} is a pre-trained sound classifier model that uses a transformer encoder architecture~\cite{vaswani2017attention}. The transformer encoding has an embedding dimension of 768, 12 layers, and 12 heads. The input raw audio is converted to a 128-dimensional log Mel filterbank using a 25ms Hamming window with 10ms slide. The spectrogram is divided into a sequence of 16$\times$16 patches with an overlap of 6 in both time and frequency. We flatten each patch to a 1D patch embedding of size 768 with a linear projection layer. A trainable positional embedding of size 768 captures the spatial structure of the 2D spectrogram. AST uses cross-modality transfer learning by using the pre-trained vision transformer (ViT) trained on ImageNet~\cite{deng2009imagenet}, assuming that the image and audio spectrogram have a similar format, which also helps to reduce computational complexity. As spectrograms are single-channel images, AST averages the 3-channel weight of ViT to make it comparable to the spectrograms. It uses cut and bi-linear interpolation to match the input positional dimension. 

Figure~\ref{fig:model architecture} illustrates the overall model architecture for fine-tuning the AST model. After the transformer layer, we add two fully connected (FC) layers of dimensions 3072 and 768 with layer normalization for normalized data distributions and faster training. A linear layer with sigmoid activation maps the audio spectrogram representation to labels for classification. We normalize the input using the training dataset to make the dataset mean and standard deviation 0 and 0.5. We fine-tune for 25 epochs on both the public and collected infant-centric datasets using two NVIDIA RTX 3090Ti and a single NVIDIA GTX 1080 Ti GPUs, respectively.
% Fine-tuning on the public dataset consists of 25 epochs of fine-tuning using two NVIDIA RTX 3090Ti GPUs and 25 epochs of fine-tuning using a single NVIDIA GTX 1080 Ti for the collected infant-centric audio.
We used a multistep learning rate (LR) scheduler with $10^{-5}$ starting rate and 0.85 decay and saved the best model for inferring the test data. We use accuracy, unweighted precision, recall, F1-scores, and Cohen's Kappa\cite{cohen1960coefficient} as evaluation metrics.

\section{Results \& Discussion}
First, we evaluate the fine-tuned AST on public data and compare against baseline algorithms. Next, we demonstrate the performance of fine-tuned AST on the infant-centric audio data collected from 22 families and evaluate our proposed training schemes using the public and collected datasets.

\subsection{Evaluation on Public Dataset}
We compare the performance of AST with one of the most recent pre-trained environmental acoustic event classification models, Whisper-AT. Whisper-AT is based on Whisper ~\cite{radford2023robust}, which is trained on large noise-conditioned speech audio to learn the noise signature inherently during training for ASR tasks. 
However, in Table~\ref{tab:whisper}, we observe that AST outperforms
Whisper-AT on popular sound classification datasets -- ESC-50 and AudioSet. Although Whisper-large~\cite{radford2023robust} is fine-tuned end-to-end to generate Whisper-AT and has 665 million parameters, it fails to outperform a fine-tuned AST with 87 million parameters when the whole model has been fine-tuned as Whisper is mainly trained on ASR tasks.

% \begin{table}[]
% \caption{Comparison of AST and whisper-AT}
% \label{tab:whisper}
% \centering
% \begin{tabularx}{0.95\columnwidth}{@{}lccc@{}}
% \toprule
%                                    & AST  & Whisper-AT & \# of Parameter\\ \midrule
% ESC-50 (Accuracy)                  & 95.6 & 91.4    &   \\
% Audioset (Mean Precision) & 48.5 & 41.5   &    \\ \bottomrule
% \end{tabularx}
% \end{table}

\begin{figure}[t]
%\vspace{-1em}
    \centering
    \includegraphics[width=.5\textwidth]{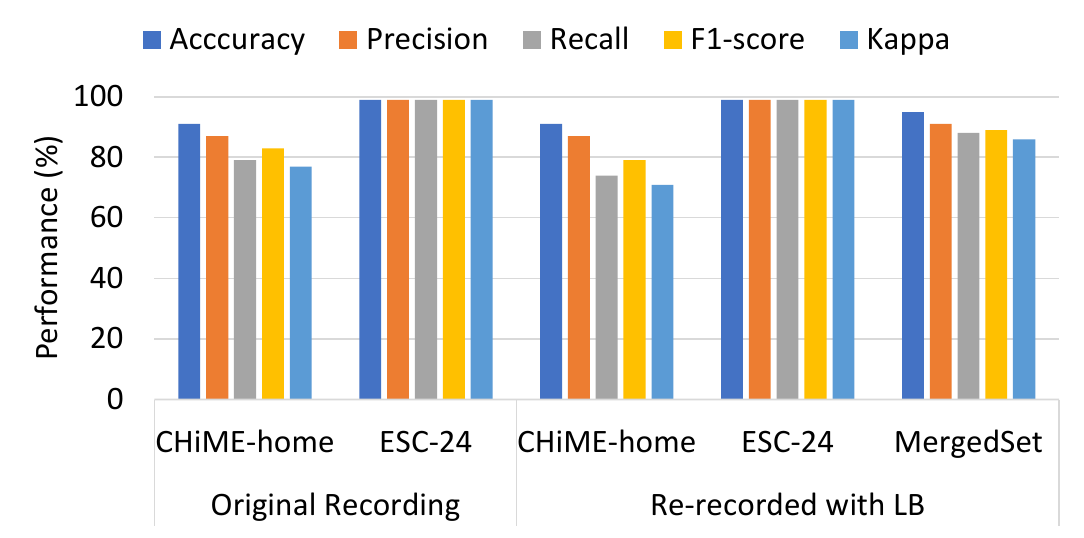}
    \vspace{-2em}
    \caption{Evaluation of AST on public dataset}
    \label{fig:public-data}
    \vspace{-2em}
\end{figure}

\begin{table}[]
\caption{Comparative Analysis of the Performance between AST and Whisper-AT.}
\vspace{-0.8em}
\label{tab:whisper}
\centering
\begin{tabularx}{0.98\columnwidth}{@{}lccc@{}}
\toprule
                                   &  \begin{tabular}[c]{@{}c@{}}ESC-50 \\ (Accuracy)\end{tabular}     & \begin{tabular}[c]{@{}c@{}}Audioset \\ (Mean Average Precision)\end{tabular} & \begin{tabular}[c]{@{}c@{}}Number of \\Parameters\end{tabular}  \\ \midrule
Whisper-AT~\cite{gong_whisperat}    & 0.91 & 0.42  &  665M   \\ 
AST~\cite{gong21b_interspeech}                & \textbf{0.96} & \textbf{0.48}  &  \textbf{87M}  \\
AST (fine-tuned)     &0.95 & 0.46  &  87M   \\ \bottomrule
\end{tabularx}
\vspace{-1em}
\end{table}
Next, we evaluate the performance of AST on two different datasets -- CHiME-home and ESC-24 (original and re-recorded with LB device) as described in Section~\ref{sec:data}. Figure~\ref{fig:public-data} shows that AST performs well with 83\% and 99\% of F1-score for CHiME-home and ESC-24, respectively.
% \begin{table}[]
% \caption{Evaluation of AST on public dataset}
% \label{tab:public-data}
% \resizebox{.49\textwidth}{!}{
% \begin{tabular}{@{}ccccccc@{}}
% \toprule
% \multicolumn{1}{l}{}                 &            & Accuracy & Precision & Recall & F1-score & Kappa \\ \midrule
% \multirow{3}{*}{Original Recording} & CHiME-home & 91        & 87        & 79     & 83       & 77    \\
%                                     & ESC-24     & 99        & 99        & 99     & 99       & 99    \\
% & CHiME-home & 91        & 87        & 74     & 79       & 71    \\ \midrule
% \multirow{2}{*}{Re-recorded with LB}                                     & ESC-24     & 99        & 99        & 99     & 99       & 99    \\
% \multicolumn{1}{l}{}                 & MergedSet  & 95        & 91        & 88     & 89       & 86    \\ \bottomrule
% \end{tabular}}
% \end{table}
To study the device-specific effect of LB on the classification, we evaluate AST on the acoustic recording of these two datasets when recorded with LB. In Figure~\ref{fig:public-data}, we observe that performance slightly degrades with a 4.8\% F1-score drop for CHiME-home. This shows that using a low-cost and lightweight recording device compared to a professional microphone has negligible impact on the classification performance. We further observe that in both recording scenarios (original publicly available data and recorded LB data), the performance is higher for ESC-24 than CHiME-home due to unbalanced training data in CHiME-home. Additionally, we get identical values for all evaluation metrics on the original ESC-24 and LB recorded ESC-24 due to a perfectly balanced dataset. Thus, we merge four public datasets to generate a balanced acoustic domestic environment dataset that reflects the soundscape around an infant at home. To understand the effect of environmental parameters, we combine data re-recorded with LB (ESC-24 and CHiME-home) and originally collected  (GTZAN and libriTTS) to create the dataset. We call this combined dataset \textbf{MergedSet}. This dataset combines all data from CHiME-home (child voice, adult speech, TV sound, percussive noise, white noise) with adult speech data from LibriSpeech-small, music data from GTZan, and percussive noise from ESC-24. 
% \note{we did not include appliance as we have enough background data to train the model}. 
We further add synthesized white noise with random Gaussian noise for mean 0 and standard deviation 1. 

Figure \ref{fig:public-data} shows that AST performs better on MergedSet than CHiME-home. MergedSet has a 12.7\% improvement in the F1-score and a 21.1\% improvement in the Cohen's Kappa score than LB recorded CHiME-home. It also outperforms the original CHiME-home data with a 7.2\% and 11.7\% improvement in the F1-score and Cohen's Kappa score, respectively.

% In this study, We merged data from public datasets to generate a domestic environment from an infant's perspective. Before merging the datasets, we evaluated the model performance on the individual data for environmental audio classifications. We also ran evaluations on the LB-recorded public data to check the model's performance on shifted domains. These results are presented in figure \ref{fig:public-data}

% As AST already trained on Audioset dataset with two million audio segments, it not only learned the features of non-human sound, but it also learned human and child voices. Thus, this model performs well on the CHiME-home and ESC-50 dataset. However, when we changed the domain by recording the same data with LB, performance on CHIME-home degraded by a little. We also find that both model works better on ESC-50 dataset than CHiME-home due to the presence of unbalanced data during training in CHIME. Thus, when generating the infant ego-centric environment by merging data from the public datasets, we used a balance sampling technique.
% We got adult speech from libespeech small, music from GTAN, percussive noise from ESC-50 and merged them with the existing data from CHiME. We also synthesize white noise using random Gaussian noise for mean 0 and standard deviation 1. With almost balanced sampling, AST performs better than unbalanced CHiME data, which is evident from figure \ref{fig:public-data}

\subsection{Evaluation on Infant LB Data in the Home}
Data collected using the LB at home are highly unbalanced and have very few samples of certain classes (e.g., 318 samples of TV and 290 samples of music) due to the uncontrolled and unscripted nature of the data collection process. As the samples are highly sparse and unbalanced for fine-tuning, we use three different training schemes to fine-tune the AST model -- (1) training on public data and evaluation on LB audio data (named \textbf{Public Data}), (2) balancing training data with resampling using LB training audio (named \textbf{Resampled Data}), and (3) training using both public and LB data while evaluating only on collected LittleBeats data (named \textbf{Mixed Data}).

% Data collected from LB home recording is highly unbalanced and very low samples in classes even for fine-tuning. We compare the performance of AST and whisper-AT model on three different training schemes: balancing training data by resampling, training on public data and evaluation on LB audio, training using both public and LB audio, and evaluation on LB audio. 
% \begin{figure}[!htb]
% \vspace{-1em}
%     \centering
%     \includegraphics[width=.5\textwidth]{figure/lb_audio_result.pdf}
%     \vspace{-2em}
%     \caption{Evaluation of AST on LB audio using three different training datasets: public data only, resampled LB audio only, LB audio and public data.}
%     \label{fig:lb-audio}
%     \vspace{-1em}
% \end{figure}
\begin{table}[]
\caption{Evaluation of AST on LB audio using three different training schemes.}
\label{tab:lb-audio}
\vspace{-1em}
\begin{tabular}{@{}cccccc@{}}
\toprule
               & Accuracy & Precision & Recall & F1-score & Kappa \\ \midrule
Public data    & 0.12        & 0.29        & 0.19     & 0.11       & 0.013   \\
Resampled data & 0.79        & 0.81        & 0.77     & 0.76       & 0.72    \\
Mixed data    & \textbf{0.86}        & \textbf{0.84}        & \textbf{0.84}     & \textbf{0.84}       & \textbf{0.83}    \\ \bottomrule
\end{tabular}
\vspace{-1em}
\end{table}
\begin{table}[t]
\caption{Comparison of different fine-tuning strategies of AST on Mixed data.}
\vspace{-0.8em}
\label{tab:fine-tune}
\resizebox{\columnwidth}{!}{%
\begin{tabular}{llllll}
\toprule
Fine-Tuning Layers                          & Accuracy & Precision & Recall & F1-score & Kappa \\
\midrule
Last two layers & 0.83      & 0.82       & 0.82    & 0.81      & 0.80  \\
Whole model     & \textbf{0.86}      & \textbf{0.84}      & \textbf{0.84}   & \textbf{0.84}     & \textbf{0.83}  \\
\bottomrule
\end{tabular}%
}
\vspace{-2em}
\end{table}
Table~\ref{tab:lb-audio} shows the performance of AST on LB home audio using the three different training datasets. The Public Data scheme, which infers on the model trained with public data only, fails to predict the sound classes of the LB home audio. When we fine-tune with LB home data only in the Resampled Data scheme, AST shows noteworthy performance in all metrics. We caution, however, that too much resampling when a small amount of data is available can lead to overfitting and poor performance in some classes. Finally, with the Mixed Data scheme where AST is fine-tuned with both LB home audio and public datasets to create a balanced dataset we observe an improvement of performance by 10.5\% in the F1-score. 

Finally, we assess how much of the AST model is required to be fine-tuned to perform well for real-world infant-centric sound classification at home. In Table \ref{tab:fine-tune}, we compare the performance between fine-tuning only the last two layers and the whole model. As AST is already trained on a large dataset, fine-tuning a few layers gives us good performance. However, fine-tuning the whole model is beneficial to get noise features and identify variability from a moving microphone. 
%Although fine-tuning a few layers performs well enough on the proposed task, fine-tuning the whole model performs better.
% We also compare AST's performance by fine-tuning only a few layers and the whole model. Although fine-tuning a few layers performs well enough on proposed task, fine-tuning the whole model provides better performance, as evident from Table \ref{tab:fine-tune}.
% \vspace{-0.2em}
\subsection{Discussion}
The performance of the AST on the original public datasets and LB device re-recorded audio of public datasets is comparable (4.8\% degradation of F1-score on re-recorded audio), as AST is already trained on Audioset \cite{gemmeke2017audio} to learn human and non-human speech well. AST also performs better on our MergedSet than LB recorded CHiME-home with a 12.7\% improvement in F1-score due to the balanced nature of the MergedSet. However, training using the balanced public dataset is not sufficient on real-world data obtained from infant-worn devices that bring unique challenges, including (a) the child's own vocalization and (b) the child's movement toward and away from other noise sources. Thus, training using Resampled Data improves the performance appreciably. Finally, the Mixed-data scheme not only improves the model's performance but also reduces the necessity of resampling data, which can result in overfitting due to too much oversampling. 
%Note that we get better performance on background noise prediction, as these data are collected from static positions rather than using baby-worn devices.

\section{Conclusion \& Future work}
In this study, we have collected, labeled, and analyzed environmental data from an infant-centric soundscape. We show that fine-tuning a large pre-trained model provides satisfactory performance when we combine publicly available data with a limited amount of infant-centric data collected in the home instead of using only public or collected audio. In the future, we aim to collect data from more families and improve the model’s performance. Valid assessments of noise soundscapes in the home and from the infant’s perspective may provide significant opportunities for early detection and intervention of infant behavioral or physiological disturbance due to noisy and unpredictable environments.

% This study collects, labels, and analyzes environmental data from an infant ego-centric soundscape. We show that fine-tuning a large pre-trained model provides satisfactory performance when we fuse publicly available data with a limited amount of collected infant ego-centric data instead of using only public or only collected audio. In the future, we aim to collect data from more families and improve the model's performance over all classes of noise with more robust validation.
% \section{Acknowledgement}
% This work was supported by funding from the National Institute on Drug Abuse (R34DA050256)

\bibliographystyle{IEEEtran}
\bibliography{reference}

% Generated by IEEEtran.bst, version: 1.14 (2015/08/26)
\begin{thebibliography}{10}
\providecommand{\url}[1]{#1}
\csname url@samestyle\endcsname
\providecommand{\newblock}{\relax}
\providecommand{\bibinfo}[2]{#2}
\providecommand{\BIBentrySTDinterwordspacing}{\spaceskip=0pt\relax}
\providecommand{\BIBentryALTinterwordstretchfactor}{4}
\providecommand{\BIBentryALTinterwordspacing}{\spaceskip=\fontdimen2\font plus
\BIBentryALTinterwordstretchfactor\fontdimen3\font minus \fontdimen4\font\relax}
\providecommand{\BIBforeignlanguage}[2]{{%
\expandafter\ifx\csname l@#1\endcsname\relax
\typeout{** WARNING: IEEEtran.bst: No hyphenation pattern has been}%
\typeout{** loaded for the language `#1'. Using the pattern for}%
\typeout{** the default language instead.}%
\else
\language=\csname l@#1\endcsname
\fi
#2}}
\providecommand{\BIBdecl}{\relax}
\BIBdecl

\bibitem{stansfeld2015health}
S.~Stansfeld and C.~Clark, ``Health effects of noise exposure in children,'' \emph{Current environmental health reports}, vol.~2, pp. 171--178, 2015.

\bibitem{li2022environmental}
A.~Li, E.~Martino, A.~Mansour, and R.~Bentley, ``Environmental noise exposure and mental health: evidence from a population-based longitudinal study,'' \emph{American journal of preventive medicine}, 2022.

\bibitem{werchan2022signal}
D.~M. Werchan, A.~Brandes-Aitken, and N.~H. Brito, ``Signal in the noise: Dimensions of predictability in the home auditory environment are associated with neurobehavioral measures of early infant sustained attention,'' \emph{Developmental psychobiology}, vol.~64, no.~7, p. e22325, 2022.

\bibitem{erickson2017influences}
L.~C. Erickson and R.~S. Newman, ``Influences of background noise on infants and children,'' \emph{Current directions in psychological science}, vol.~26, no.~5, pp. 451--457, 2017.

\bibitem{wass2019influences}
S.~V. Wass, C.~G. Smith, K.~R. Daubney, Z.~M. Suata, K.~Clackson, A.~Begum, and F.~U. Mirza, ``Influences of environmental stressors on autonomic function in 12-month-old infants: Understanding early common pathways to atypical emotion regulation and cognitive performance,'' \emph{Journal of Child Psychology and Psychiatry}, vol.~60, 2019.

\bibitem{xu2009reliability}
D.~Xu, U.~Yapanel, and S.~Gray, ``Reliability of the lena language environment analysis system in young children’s natural home environment,'' \emph{Boulder, CO: Lena Foundation}, pp. 1--16, 2009.

\bibitem{bulgarelli2020look}
F.~Bulgarelli and E.~Bergelson, ``Look who’s talking: A comparison of automated and human-generated speaker tags in naturalistic day-long recordings,'' \emph{Behavior Research Methods}, vol.~52, 2020.

\bibitem{khante5auditory}
P.~Khante, E.~Thomaz, and K.~de~Barbaro, ``Auditory chaos classification in real-world environments,'' \emph{Frontiers in Digital Health}, vol.~5.

\bibitem{li23e_interspeech}
J.~Li, M.~Hasegawa-Johnson, and N.~L. McElwain, ``{Towards Robust Family-Infant Audio Analysis Based on Unsupervised Pretraining of Wav2vec 2.0 on Large-Scale Unlabeled Family Audio},'' in \emph{Proc. INTERSPEECH 2023}, 2023, pp. 1035--1039.

\bibitem{gong21b_interspeech}
Y.~Gong, Y.-A. Chung, and J.~Glass, ``{AST: Audio Spectrogram Transformer},'' in \emph{Proc. Interspeech 2021}, 2021, pp. 571--575.

\bibitem{gemmeke2017audio}
J.~F. Gemmeke, D.~P. Ellis, D.~Freedman, A.~Jansen, W.~Lawrence, R.~C. Moore, M.~Plakal, and M.~Ritter, ``Audio set: An ontology and human-labeled dataset for audio events,'' in \emph{2017 IEEE international conference on acoustics, speech and signal processing (ICASSP)}.\hskip 1em plus 0.5em minus 0.4em\relax IEEE, 2017.

\bibitem{piczak2015esc}
K.~J. Piczak, ``Esc: Dataset for environmental sound classification,'' in \emph{Proceedings of theACM international conference on Multimedia}, 2015.

\bibitem{guzhov2021esresnet}
A.~Guzhov, F.~Raue, J.~Hees, and A.~Dengel, ``Esresnet: Environmental sound classification based on visual domain models,'' in \emph{2020 25th International Conference on Pattern Recognition (ICPR)}.\hskip 1em plus 0.5em minus 0.4em\relax IEEE, 2021.

\bibitem{tokozume2017learning}
Y.~Tokozume and T.~Harada, ``Learning environmental sounds with end-to-end convolutional neural network,'' in \emph{2017 IEEE international conference on acoustics, speech and signal processing (ICASSP)}.

\bibitem{zhu2018environmental}
B.~Zhu, K.~Xu, D.~Wang, L.~Zhang, B.~Li, and Y.~Peng, ``Environmental sound classification based on multi-temporal resolution convolutional neural network combining with multi-level features,'' in \emph{Advances in Multimedia Information Processing-Pacific-Rim Conference on Multimedia, Hefei, China}.\hskip 1em plus 0.5em minus 0.4em\relax Springer, 2018.

\bibitem{touvron2021training}
H.~Touvron, M.~Cord, M.~Douze, F.~Massa, A.~Sablayrolles, and H.~J{\'e}gou, ``Training data-efficient image transformers \& distillation through attention,'' in \emph{International conference on machine learning}.\hskip 1em plus 0.5em minus 0.4em\relax PMLR, 2021, pp. 10\,347--10\,357.

\bibitem{gong_whisperat}
Y.~Gong, S.~Khurana, L.~Karlinsky, and J.~Glass, ``Whisper-at: Noise-robust automatic speech recognizers are also strong audio event taggers,'' in \emph{Proc. Interspeech 2023}, 2023.

\bibitem{deng2009imagenet}
J.~Deng, W.~Dong, R.~Socher, L.-J. Li, K.~Li, and L.~Fei-Fei, ``Imagenet: A large-scale hierarchical image database,'' in \emph{2009 IEEE conference on computer vision and pattern recognition}.\hskip 1em plus 0.5em minus 0.4em\relax Ieee, 2009, pp. 248--255.

\bibitem{radford2023robust}
A.~Radford, J.~W. Kim, T.~Xu, G.~Brockman, C.~McLeavey, and I.~Sutskever, ``Robust speech recognition via large-scale weak supervision,'' in \emph{International Conference on Machine Learning}, 2023.

\bibitem{nancy2023}
N.~McElwain, M.~Fisher, C.~Nebeker, J.~Bodway, B.~Islam, and M.~Hasegawa-Johnson, ``Evaluating users’ experiences of a child multimodal wearable device: A mixed methods approach (in press),'' \emph{JMIR Human Factors}, 05 2023.

\bibitem{foster2015chime}
P.~Foster, S.~Sigtia, S.~Krstulovic, J.~Barker, and M.~D. Plumbley, ``Chime-home: A dataset for sound source recognition in a domestic environment,'' in \emph{2015 IEEE Workshop on Applications of Signal Processing to Audio and Acoustics (WASPAA)}.\hskip 1em plus 0.5em minus 0.4em\relax IEEE, 2015, pp. 1--5.

\bibitem{font2013freesound}
F.~Font, G.~Roma, and X.~Serra, ``Freesound technical demo,'' in \emph{Proceedings of the ACM international conference on Multimedia}, 2013.

\bibitem{zen19_interspeech}
H.~Zen, V.~Dang, R.~Clark, Y.~Zhang, R.~J. Weiss, Y.~Jia, Z.~Chen, and Y.~Wu, ``{LibriTTS: A Corpus Derived from LibriSpeech for Text-to-Speech},'' in \emph{Proc. Interspeech 2019}, 2019, pp. 1526--1530.

\bibitem{1021072}
G.~Tzanetakis and P.~Cook, ``Musical genre classification of audio signals,'' \emph{IEEE Transactions on Speech and Audio Processing}, 2002.

\bibitem{sohn1999statistical}
J.~Sohn, N.~S. Kim, and W.~Sung, ``A statistical model-based voice activity detection,'' \emph{IEEE signal processing letters}, 1999.

\bibitem{boersma2007praat}
P.~Boersma, ``Praat: doing phonetics by computer,'' \emph{http://www. praat. org/}, 2007.

\bibitem{mcfee2015librosa}
B.~McFee, C.~Raffel, D.~Liang, D.~P. Ellis, M.~McVicar, E.~Battenberg, and O.~Nieto, ``librosa: Audio and music signal analysis in python,'' in \emph{Proceedings of the 14th python in science conference}, vol.~8, 2015.

\bibitem{wei2020comparison}
S.~Wei, S.~Zou, F.~Liao \emph{et~al.}, ``A comparison on data augmentation methods based on deep learning for audio classification,'' in \emph{Journal of physics: Conference series}, vol. 1453, no.~1.\hskip 1em plus 0.5em minus 0.4em\relax IOP Publishing, 2020.

\bibitem{vaswani2017attention}
A.~Vaswani, N.~Shazeer, N.~Parmar, J.~Uszkoreit, L.~Jones, A.~N. Gomez, {\L}.~Kaiser, and I.~Polosukhin, ``Attention is all you need,'' \emph{Advances in neural information processing systems}, vol.~30, 2017.

\bibitem{cohen1960coefficient}
J.~Cohen, ``A coefficient of agreement for nominal scales,'' \emph{Educational and psychological measurement}, vol.~20, no.~1, pp. 37--46, 1960.

\end{thebibliography}
\end{document}